# $g_{a_0\rho\gamma}$ and $g_{f_0\rho\gamma}$ coupling constants in three point QCD sum rules

C. Aydın*, M. Bayar and A. H. Yılmaz[†]

*Physics Department, Karadeniz Technical University, 61080, Trabzon, Turkey*

[*]coskun@ktu.edu.tr
[†]hakany@ktu.edu.tr

The coupling constant of $a_0 \to \rho\gamma$ and $f_0 \to \rho\gamma$ decays are calculated using 3-point QCD sum rules. We estimate the coupling constant $g_{a_0\rho\gamma}$ and $g_{f_0\rho\gamma}$ which are an essential ingredient in the analysis of physical processes involving isoscalar $f_0(980)$ and isovector $a_0(980)$ mesons.

I. **Introduction**

One of the very important goals of studying nuclear physics is to understand the behavior of hadrons and hadronic interactions on the basis of the quantum chromodynamics (QCD). One of the powerful tools for this aim is the QCD sum rule invented by Shifman, Vainshtein, and Zakharov [1] provides us with a way to relate the physical quantities of the hadrons to the matrix elements of the quark-gluon composite operators by means of the operator product expansion (OPE) [1, 2]. The field of application of the sum rules has been extended remarkably since 1980's. The QCD sum rule method has been utilized to analyze many hadronic properties, and it yields an effective framework to investigate the hadronic observables such as decay constants and form factors within the nonperturbative contributions proprotional to the quark and gluon condensates [3].

With increasing experimental information about the different members of the meson spectrum it becomes very important to develop a consistent understanding of the observed mesons from a theoretical point of view. For the low-lying pseudoscalar, vector, and tensor mesons this has been done quite successfully within the framework of the simple quark model assuming the mesons to be quark-antiquark ($q\bar{q}$) states grouped together into nonets. The decay channels of $a_0(980)$ and $f_0(980)$ mesons can be analyzed in the context of QCD sum rules.

The flavor SU(3) forms approximate global symmetry of hadron spectrum according to which mesons are classified as bound states of a quark and antiquark ($q\bar{q}$) and they are placed in nonet representations of SU(3) group. However, whether light scalar mesons form a scalar nonet is still an open question. From the experimental point of view, the isoscalar $f_0(980)$ and isovector $a_0(980)$ are well established, but the nature and the quark substructure of these scalar mesons, the question whether they are conventional $q\bar{q}$ states, $K\bar{K}$ molecules, or multiquark exotic $q^2\bar{q}^2$ states has been a subject of controversy. Understanding the nature and the quark substructure of the scalar mesons is still an open problem in hadron physics.

Radiative transitions between pseudoscalar (P) and vector (V) mesons have been an important subject in low-energy hadron physics for more than three decades. These transitions have been regarded as phenomenological quark models, potential models, bag models, and effective Lagrangian methods [4,5]. Among the characteristics of the electromagnetic interaction processes $g_{VP\gamma}$ coupling constant plays one of the most important roles, since they determine the strength of the hadron interactions. Low-energy hadron interactions are governed by nonperturbative QCD so that it is very difficult to get the numerical values of the coupling constants from the first principles. For that reason a semiphenomenological method of QCD sum rules can be used, which nowadays is the standart tool for studying of various characteristics of hadron interactions. On the other hand, vector meson-pseudoscalar meson-photon VPγ–vertex also plays a role in photoproduction reactions of vector mesons on nucleons. It should be notable that in these decays (V→Pγ) the four-momentum of the pseudoscalar meson P is time-like, $p'^2 > 0$, whereas in the pseudoscalar exchange amplitude contributing to the photoproduction of vector mesons it is space-like $p'^2 < 0$. Therefore, it is of interest to study the effective coupling constant $g_{VP\gamma}$ from another point of view as well.

In this work, we studied $a_0 \to \rho\gamma$ and $f_0 \to \rho\gamma$ decay in the framework of three-point QCD sum rules and we obtained the coupling constant $g_{a_0\rho\gamma}$ and $g_{f_0\rho\gamma}$.

## II. Calculation

According to the general strategy of QCD sum rules method, the coupling constants can be calculated by equating the representations of a suitable correlator calculated in terms of hadronic and quark-gluon degrees of freedom. In order to do this we consider the following correlation function by using the appropriately chosen currents

$$\Pi_{\mu\nu}(p,p') = \int d^4x\, d^4y\, e^{ip'\cdot y} e^{-ip\cdot x} <0|T\{J_\mu^\gamma(0) J_{f_0}(x) J^\omega(y)\}|0> \quad (1)$$

We choose the interpolating current for the ρ and S mesons as $j_\mu^\rho = \frac{1}{2}(\bar{u}\gamma_\mu^a u^a - \bar{d}\gamma_\mu^a d^a)$, and $J_S$ (for $a_0$, $J_{a_0} = \frac{1}{2}[\bar{u}^b u^b - \bar{d}^b d^b]$ and for $f_0$, $J_{f_0} = \frac{1}{\sqrt{2}}[\bar{u}^b u^b + \bar{d}^b d^b]\sin\theta + \bar{s}s\cos\theta$ [6] respectively. ρ-meson consist of $u$ and $d$-quarks. $J_\mu^\gamma = e_u(\bar{u}\gamma_\mu u) + e_d(\bar{d}\gamma_\mu d)$ is the electromagnetic current with $e_u$ and $e_d$ being the quark charges.

The theoretical part of the sum rule in terms of the quark-gluon degrees of freedom for the coupling constant $g_{a_0\rho\gamma}$ and $g_{f_0\rho\gamma}$ are calculated by considering the perturbative contribution and the power corrections from operators of different dimensions to the three-point correlation function $\Pi_{\mu\nu}$. For the perturbative contribution we study the lowest order bare-loop diagram. Moreover, the power corrections from the operators of different dimensions $<\bar{q}q>$, $<\bar{q}\eta.Gq>$, and $<(\bar{q}q)^2>$ are considered in this work. Since it is estimated to be negligible for light quark systems, we did not consider the gluon condensate contribution proportional to $<G^2>$. We perform the calculations of the power corrections in the fixed point gauge [7]. We also work in the limit $m_q = 0$ and in this limit the perturbative bare-loop diagram does not make any contribution. In fact, by considering this limit only operators of dimensions d=3 and d=5 make contributions which are proportional to $<\bar{q}q>$ and $<\bar{q}\eta.Gq>$, respectively. The relevant Feynman diagrams for power corrections are given in Fig 1.

On the other hand, in order to calculate the phenomenological part of the sum rule in terms of hadronic degrees of freedom, a double dispersion relation satisfied by the vertex function $\Pi_{\mu\nu}$ is considered [1, 2, 8]:

$$\Pi_{\mu\nu}(p,p') = \frac{1}{\pi^2}\int ds_1 \int ds_2 \frac{\rho_{\mu\nu}(s_1,s_2)}{(p^2-s_1)(p'^2-s_2)} \qquad (2)$$

where we ignore possible substruction terms since they will not make any contributions after Borel transformation. For our purpose we choose the vector and pseudoscalar channels and saturating this dispersion relation by the lowest lying meson states in these channels the physical part of the sum rule is obtained as

$$\Pi_{\mu\nu}(p,p') = \frac{<0|J^\rho|S><S(p)|J_\mu^\gamma|\rho(p')><\rho|J_s|0>}{(p^2-m_S^2)(p'^2-m_\rho^2)} + ..., \qquad (3)$$

where the contributions from the higher states and the continuum are given by dots. The overlap amplitudes for vector and pseudoscalar mesons are $<0|J_\mu^\rho|\rho>=\lambda_\rho \varepsilon_\mu^\rho$, where $\varepsilon_\mu^\rho$ is the polarization vector of the vector meson and $<S|J_s|0>=\lambda_S$, respectively, where $S=a_0$ or $f_0$. The matrix element of the electromagnetic current is given by

$$<S(p)|J_\mu^\gamma|\rho(p')>=-i\frac{e}{m_\rho}g_{S\rho\gamma}K(q^2)(p.qu_\mu - u.qp_\mu) \qquad (4)$$

where $q = p - p'$ and $K(q^2)$ is a form factor with $K(0)=1$. This matrix element defines the coupling constant $g_{SV\gamma}$ by means of the effective Lagrangian

$$\mathcal{L} = \frac{e}{m_\rho}g_{S\rho\gamma}\partial_\mu\rho_\nu(\partial_\nu A_\beta - \partial_{\beta}A_\nu)S \qquad (5)$$

describing the $S\rho\gamma$ – vertex [9].

We perform the calculations of the power corrections in the fixed-point gauge $x_\mu A^\mu = 0$. The general forms of the contributions corresponding to Feynman diagrams are derived with respect to their dimensions. For the first diagrams in Fig. 1a, there are two contributions with different dimensions for $d \leq 5$ as

$$F_2(3d) = -N_c \langle \bar{\psi}\psi \rangle \frac{1}{4} Tr\left[ \Gamma_1 \frac{1}{\not{p}' - m_q} \Gamma_2 \frac{1}{\not{p} - m_q} \Gamma_3 \right] \quad (6)$$

$$F_2(5d) = -\frac{N_c}{2} \langle 0 | \bar{\psi}_\alpha^a(0) \nabla_\lambda \nabla_{\lambda'} \psi_\beta^a(x) | 0 \rangle x_\lambda x'_{\lambda'} Tr\left[ \Gamma_1 \frac{1}{\not{p}' - m_q} \Gamma_2 \frac{1}{\not{p} - m_q} \Gamma_3 \right]_{\alpha\beta} \quad (7)$$

$$F_2(5d) = -\frac{N_c}{2} \frac{m_q^2}{16} \langle \bar{\psi}\psi \rangle \frac{\partial}{\partial p_\lambda} \frac{\partial}{\partial p_\lambda} Tr\left[ \Gamma_1 \frac{1}{\not{p}' - m_q} \Gamma_2 \frac{1}{\not{p} - m_q} \Gamma_3 \right]$$

$$+ N_c \frac{1}{64} \langle \bar{\psi} g G^a_{\lambda\lambda'}(\lambda^a/2) \sigma_{\lambda\lambda'} \psi \rangle \frac{\partial}{\partial p_\lambda} \frac{\partial}{\partial p_\lambda} Tr\left[ \Gamma_1 \frac{1}{\not{p}' - m_q} \Gamma_2 \frac{1}{\not{p} - m_q} \Gamma_3 \right] \quad (8)$$

For the second and third diagrams in Fig. 1b and 1c, there are contributions with dimensions for $d \leq 5$ as

$$F_3(5d) = -iN_c \frac{g}{192} \langle \bar{\psi} G^a_{\lambda\rho} \sigma_{\lambda\rho}(0) \frac{\lambda^c}{2} \psi \rangle \frac{\partial}{\partial k_\lambda} Tr\left[ \Gamma_1 \frac{1}{\not{p}' - m_q} \gamma_\rho \frac{1}{\not{p}' - \not{k} - m_q} \Gamma_2 \frac{1}{\not{p} - m_q} \Gamma_3 \sigma_{\lambda\rho} \right]_{k=0} \quad (9)$$

$$F'_3(5d) = -iN_c \frac{g}{192} \langle \bar{\psi} G^a_{\lambda\rho} \sigma_{\lambda\rho}(0) \frac{\lambda^c}{2} \psi \rangle \frac{\partial}{\partial k_\lambda} Tr\left[ \Gamma_1 \frac{1}{\not{p}' - m_q} \Gamma_2 \frac{1}{\not{p} - \not{k} - m_q} \gamma_\rho \frac{1}{\not{p} - m_q} \Gamma_3 \sigma_{\lambda\rho} \right]_{k=0} \quad (10)$$

For $a_0(f_0) \to \rho\gamma$ decay the vertex functions are $\Gamma_1 = -\frac{i}{2}\gamma_\nu$, $\Gamma_2 = -ie_q\gamma_\mu$, and $\Gamma_3 = -\frac{i}{2}$ for $a_0$, $\Gamma_3 = -\frac{i}{\sqrt{2}}\sin\theta$ for $f_0$. Using these vertex functions in Eqs.(6-10) we then get the nonvanishing contributions from power corrections to the correlation function as

$$\Pi_{\mu\nu} = CN_c \langle \bar{\psi}\psi \rangle [-\frac{1}{p'^2 p^2}$$

$$+ \frac{m_0^2}{4}(\frac{1}{p'^4 p^2} + \frac{1}{p'^2 p^4} - \frac{1}{6}\frac{1}{p'^4 p^2} - \frac{1}{2}\frac{1}{p'^2 p^4})](p_\nu p'_\mu - p \cdot p' g_{\mu\nu}) \quad (11)$$

where $C = \frac{1}{4}$ for $a_0$ and $C = \frac{1}{2\sqrt{2}}\sin\theta$ for $f_0$.

The lowest order perturbative quark loop diagrams do not make any contributions.

The structure $(p_\nu p'_\mu - p \cdot p' g_{\mu\nu})$ is chosen to compere to theoretical and phenomenological parts and to obtain the coupling constant $g_{S\rho\gamma}$. We then find theoretical part of the invariant function $T_1$ for $S \to \rho\gamma$ decay as

$$T_1 = C N_c \langle \bar\psi\psi \rangle \left( -\frac{1}{p'^2 p^2} + \frac{5}{24} m_0^2 \frac{1}{p'^4 p^2} + \frac{1}{8} m_0^2 \frac{1}{p'^2 p^4} \right) \tag{12}$$

After performing the double Borel transform with respect to the variables $Q^2 = -p^2$ and $Q'^2 = -p'^2$, and by considering the gauge-invariant structure $(p_\nu p'_\mu - p \cdot p' g_{\mu\nu})$, we obtain the sum rule for the coupling constants,

$$g_{f_0\rho\gamma} = (e_u - e_q) \frac{3 m_\rho}{2\sqrt{2} \lambda_\rho \lambda_{f_0}} e^{m_{f_0}^2 / M_1^2} e^{m_\rho^2 / M_2^2} \langle \bar u u \rangle \left( -3 - \frac{3}{8} \frac{m_0^2}{M_1^2} - \frac{5}{8} \frac{m_0^2}{M_2^2} \right) \sin\theta \tag{13}$$

and

$$g_{a_0\rho\gamma} = (e_u + e_q) \frac{3 m_\rho}{4 \lambda_\rho \lambda_{a_0}} e^{m_{a_0}^2 / M_1^2} e^{m_\rho^2 / M_2^2} \langle \bar u u \rangle \left( -3 - \frac{3}{8} \frac{m_0^2}{M_1^2} - \frac{5}{8} \frac{m_0^2}{M_2^2} \right) \tag{14}$$

where $M_1^2$ and $M_2^2$ are Borel masses corresponding to $a_0$ ($f_0$) and $\rho$ mesons, respectively.

### III. Numerical Results and Discussion

For the numerical evaluation of sum rule we use the values $\langle \bar u u \rangle = -0.014 \text{ GeV}^3$, $m_{f_0} = 0.98 \text{ GeV}$, $m_{a_0} = 0.98 \text{ GeV}$, $\lambda_{f_0} = 0.18 \pm 0.02 \text{ GeV}^2$ [10], $\lambda_{a_0} = 0.21 \pm 0.05 \text{ GeV}^2$ [11], $m_\rho = 0.770 \text{ GeV}$. We note that neglecting the electron mass the $e^+ e^-$ decay width of $\rho$ meson is given as $\Gamma(\rho \to e^+ e^-) = \frac{4\pi\alpha^2}{3} \left( \frac{\lambda_\rho}{3} \right)^2$. Then using the value from the experimental leptonic decay width $\Gamma(\rho \to e^+ e^-) = 7.02 \pm 0.11 \text{ keV}$ for $\rho$ [10, 12], we obtain the value $\lambda_\rho = (0.17 \pm 0.03) \text{ GeV}^2$ for the overlap amplitude $\rho$ meson. In order to examine the dependence of $g_{a_0\rho\gamma}$ and $g_{f_0\rho\gamma}$ on the Borel masses $M_1^2$ and $M_2^2$, we choose $M_1^2 = 1.2$, $1.3$ and $1.4\,\text{GeV}^2$ for $g_{a_0\rho\gamma}$ and $M_1^2 = 0.7$, $0.8$ and $0.9\,\text{GeV}^2$ for $g_{f_0\rho\gamma}$. Since the Borel mass $M^2$ is an auxiliary parameter and the physical quantitites should not depend on it, one must look for the region where $g_{a_0\rho\gamma}$ and $g_{f_0\rho\gamma}$ are practically independent of $M^2$. We first determined that this condition is satisfied in the interval $1.0\,\text{GeV}^2 \leq M_2^2 \leq 1.6\,\text{GeV}^2$ for $g_{a_0\rho\gamma}$ and $0.8\,\text{GeV}^2 \leq M_2^2 \leq 1.2\,\text{GeV}^2$ for $g_{f_0\rho\gamma}$. The variation of the coupling constant $g_{a_0\rho\gamma}$ as a function of Borel parameters $M_2^2$ for different values of $M_1^2$ are shown in Figs. 2 and 4 while for $g_{f_0\rho\gamma}$ in Figs. 3 and 5. Examination of these figures show that the sum rule is rather stable with these reasonable variations of $M_1^2$ and $M_2^2$. We then choose the middle value $M_2^2 = 1.3\,\text{GeV}^2$ for the Borel parameter in its interval of variation and obtain the coupling constant of $g_{a_0\rho\gamma}$ as between $g_{a_0\rho\gamma} = 0.96 \pm 0.44$ and $g_{a_0\rho\gamma} = 0.85 \pm 0.38$ and also for $g_{f_0\rho\gamma}$ we find the coupling constant $g_{f_0\rho\gamma}$ for the middle value $M_2^2 = 1.0\,\text{GeV}^2$ at $\theta = 30'$ as between $g_{f_0\rho\gamma} = 1.75 \pm 0.53$ and $g_{f_0\rho\gamma} = 1.12 \pm 0.34$, where only the error arising from the numerical analysis of the sum rule is considered. When we use the mixing angle for $\theta = -68'$ [14] we have $g_{f_0\rho\gamma}$ as $3.24 \pm 0.97 \leq |g_{f_0\rho\gamma}| \leq 2.07 \pm 0.62$ at $M_2^2 = 1.0\,\text{GeV}^2$. In the previous work [13] coupling constants of $g_{a_0\rho\gamma}$ and $g_{f_0\rho\gamma}$ were found as $g_{a_0\rho\gamma} = 0.85 \pm 0.36$ and $g_{f_0\rho\gamma} = 1.97 \pm 0.57$ in the framework of light-cone QCD sum rules. The coupling constant $g_{a_0\rho\gamma}$ was also calculated [10] as $2.0 \pm 0.50$ and $1.30 \pm 0.30$ in QCD sum rules.

**Acknowledgments**

This work partly supported by the Research Fund of Karadeniz Technical University, under grant contact no 2002.111.001.2.

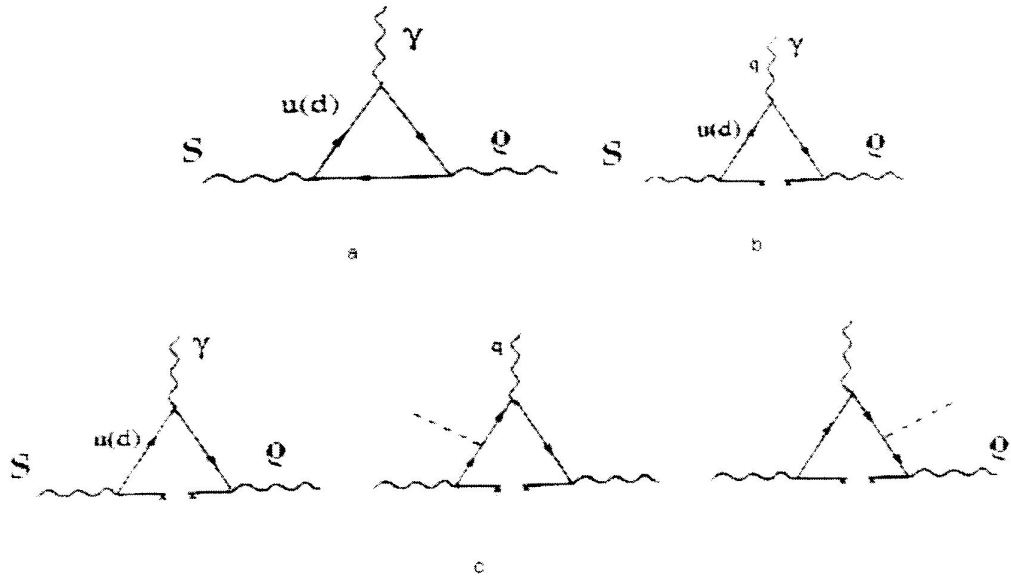

Figure 1. Feynman diagrams for the $S\rho\gamma$-vertex: a) bare loop diagram, b) d=3 operator corrections, and c) d=5 operator corrections. The dotted lines denote gluons.

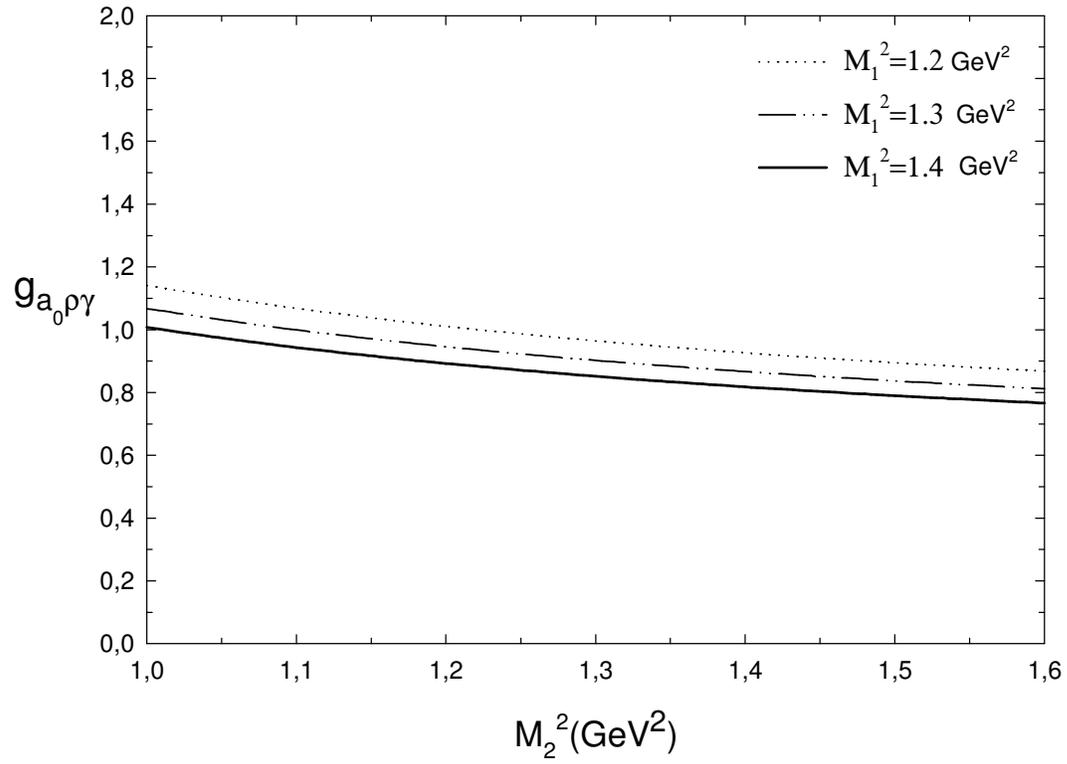

Figure 2. The coupling constant $g_{a_0\rho\gamma}$ as a function of the Borel parameter $M_1^2$ for different values of $M_1^2$.

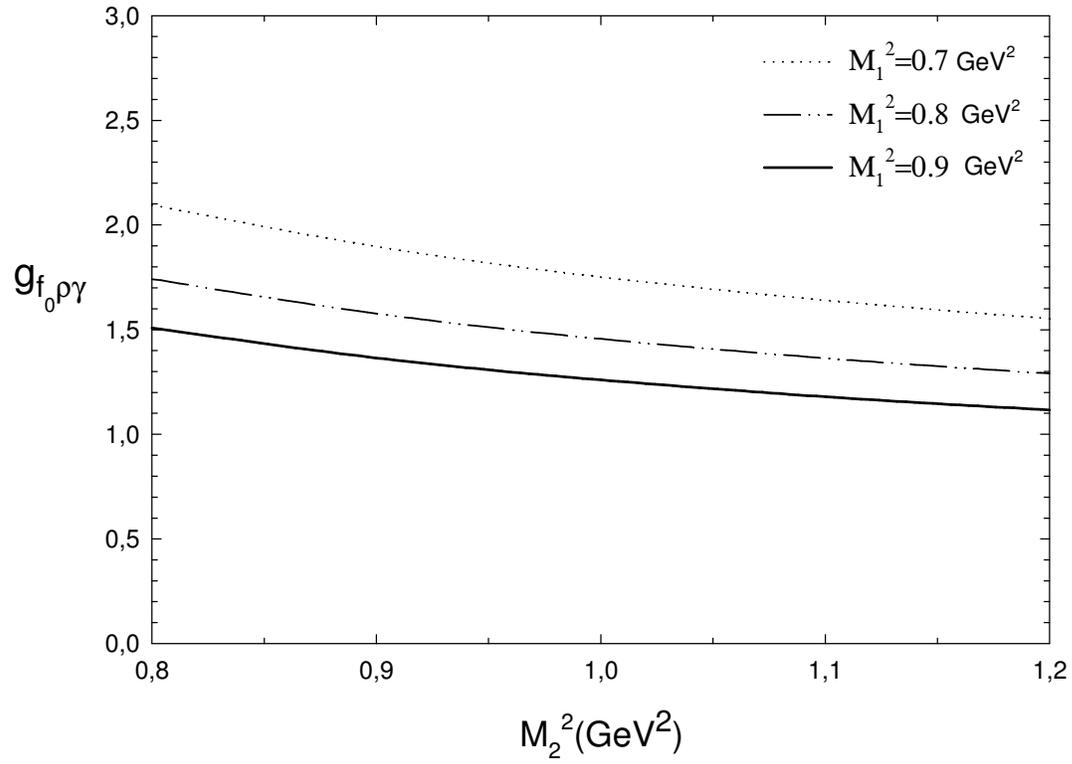

Figure 3. The coupling constant $g_{a_0\rho\gamma}$ as a function of the Borel parameter $M_1^2$ for different values of $M_1^2$ at $\theta = 30'$.

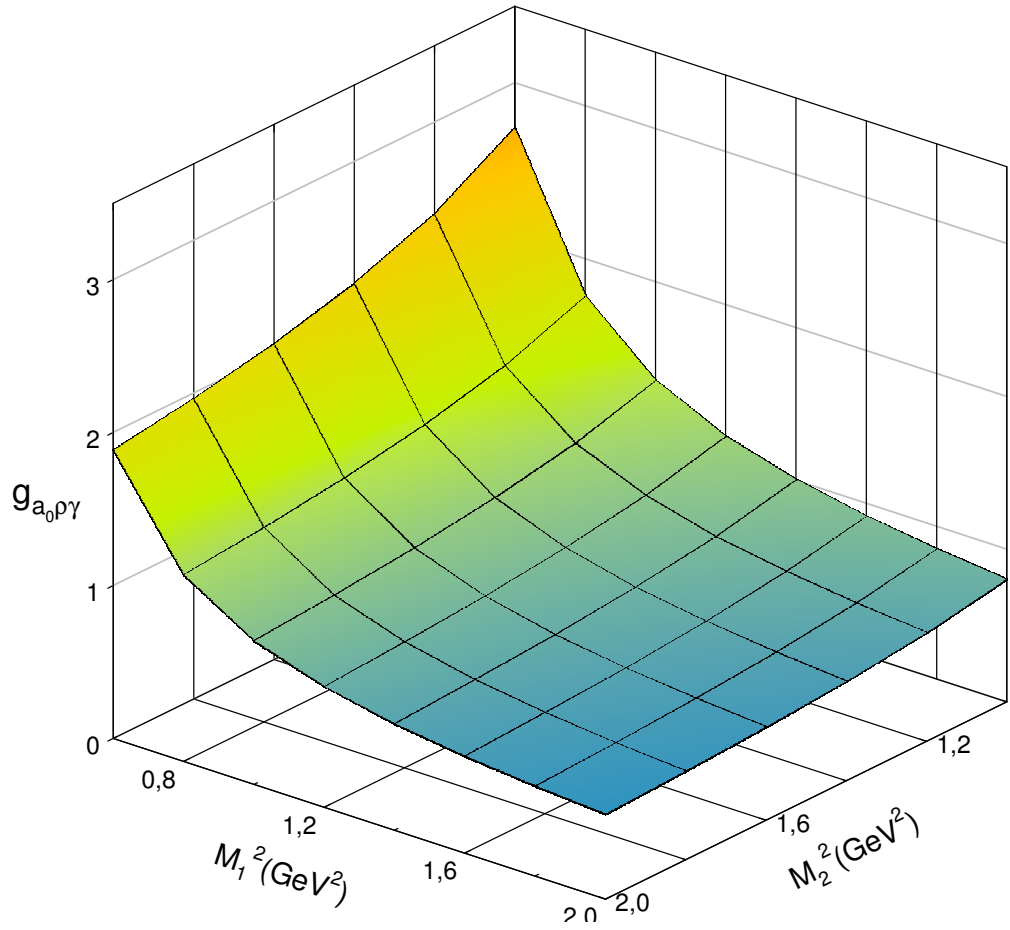

Figure 4. Coupling $g_{a_0\rho\gamma}$ as function of the Borel parameters $M_2^2$ for different values of $M_1^2$.

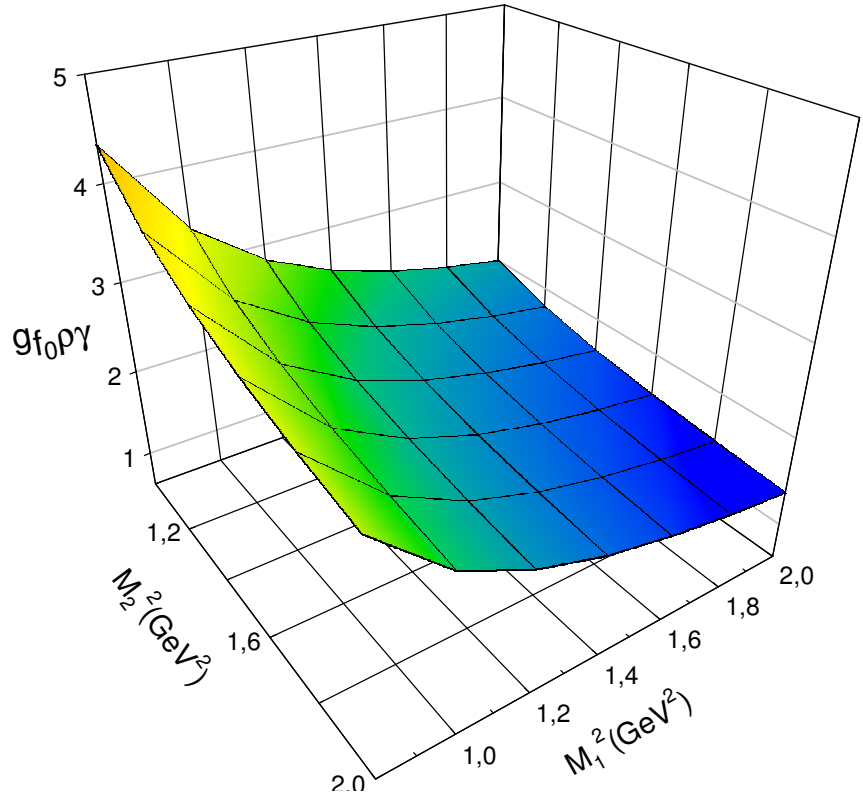

Figure 5. Coupling $g_{f_0\rho\gamma}$ as function of the Borel parameters $M_2^2$ for different values of $M_1^2$.